\documentclass{epl}
\usepackage{amssymb,graphicx,psfrag,amsmath}

\title{Optical conductivity of superconducting Sr$_2$RuO$_4$}
\shorttitle{Optical conductivity of superconducting Sr$_2$RuO$_4$}
\author{Bal\'azs D\'ora\inst{1} \and Kazumi Maki\inst{2} 
\and Attila Virosztek\inst{3,4}}
\institute{
\inst{1} The Abdus Salam ICTP, Strada Costiera 11, Trieste, I-34014, Italy\\
\inst{2} Department of Physics and Astronomy, University of Southern California, Los Angeles
CA 90089-0484, USA \\
\inst{3} Department of Physics, Budapest University of Technology and Economics,
 H-1521 Budapest, Hungary \\
\inst{4} Research Institute for Solid State Physics and Optics, P.O.Box 49,
H-1525 Budapest, Hungary}
\shortauthor{Bal\'azs D\'ora \etal}


\date{}

\begin{document}
\maketitle
\begin{abstract}
We compute the optical conductivity of 2D f-wave superconductors and also
of the multigap model proposed recently by Zhitomirsky and Rice at $T=0$K in the Born
limit. The presence
of interband impurity scattering   was  found   to play an important role:
the contributions 
from the two bands mix up, and new structures
are seen in the tunneling density of states and in the optical spectrum as
well, corresponding to interband transitions.
 This will provide a sensitive test in selecting
the competing models for the triplet superconductivity in Sr$_2$RuO$_4$.

\end{abstract}

\section{Introduction}

The discovery of superconductivity in Sr$_2$RuO$_4$ in 1994 has generated
much interest since it is a perovskite with the same crystal structure as
La$_2$CuO$_4$\cite{maeno}. From analogy with superfluid $^3$He Rice and Sigrist\cite{rs,mrs}
proposed the spin triplet p-wave superconductor with order parameter
\begin{equation}
{\bf \Delta}({\bf k})={\bf d}\Delta_p(T)(\hat k_x\pm i\hat k_y)
\label{elso}
\end{equation}
Here $\bf d$ is the spin vector parallel to $z$.

The triplet pairing has been confirmed by the presence of spontaneous
magnetization seen by $\mu$SR\cite{luke}, which revealed that the
superconducting state breaks the time-reversal symmetry, and a flat $^{17}$O Knight shift seen by
NMR\cite{ishida2} showed no change in the spin susceptibility when passing
through the superconducting transition temperature. 
However the large density of states ($N(0)$) in the gapless region was
somewhat surprising\cite{nishizaki2}, since eq. (\ref{elso}) has the full energy gap\cite{p-wavemaki}.

Furthermore, as sample quality improved, the characteristics of nodal
superconductors became clearly visible in the specific heat ($\propto
T^2$)\cite{nishizaki},
the magnetic penetration depth ($\propto T^2$)\cite{bonalde}, NMR relaxation rate
($\propto T^3$)\cite{ishida}, the thermal conductivity ($\propto T^2$)\cite{tanatar1,tanatar2} and the
ultrasonic
attenuation\cite{lupien}.
Therefore a variety of 2D f-wave models have been proposed\cite{miyake,graf,dahm}. Since all these
f-wave models have the same quasiparticle density of states, the specific
heat and the magnetic penetration depth etc. are described equally well by
any of these 2D f-wave models\cite{dahm}.
On the other hand, the angular dependent magnetothermal conductivity data
by Izawa et al.\cite{izawa} can exclude all 2D f-wave models with nodes lying in the
$a-b$ plane\cite{dahm,won}.

Therefore it singles out the most consistent f-wave as
\begin{equation}
{\bf \Delta}({\bf k})={\bf d}\Delta_f(T) (\hat k_x\pm i\hat k_y)\cos(ck_z).
\end{equation}
The nodal lines are horizontal. Also the pair correlation takes the maximum
value for the relative separation of the pair at $\pm c$.
However in quasi-two dimensional systems with paramagnon (i.e. the
ferromagnetic fluctuation), the ground state of 
p-wave superconductors is given by eq. (\ref{elso})\cite{mrs}.
This is why Zhitomirsky and Rice (ZR)\cite{ZR} proposed the
multigap model: there is p-wave superconductor attached the active $\gamma$
band, which induces another f-wave like order parameter in the
passive $\alpha+\beta$ bands through Cooper pair scattering given by
\begin{equation}
{\bf \Delta}_2({\bf k})={\bf d}\Delta_2(T) (\hat k_x\pm i\hat k_y)\cos(ck_z/2).
\label{harom}
\end{equation}
Since ${\bf \Delta}_2(\bf k)$ as given in eq. \ref{harom} gives the same
quasiparticle density of states as a d-wave superconductor\cite{d-wavemaki}, it is not
difficult to reproduce the specific heat and the magnetic penetration depth
within the multigap model as was shown recently by Kusunose and Sigrist\cite{kusunose}.

\begin{figure}[h!]
\centering
\includegraphics[width=2cm,height=4cm]{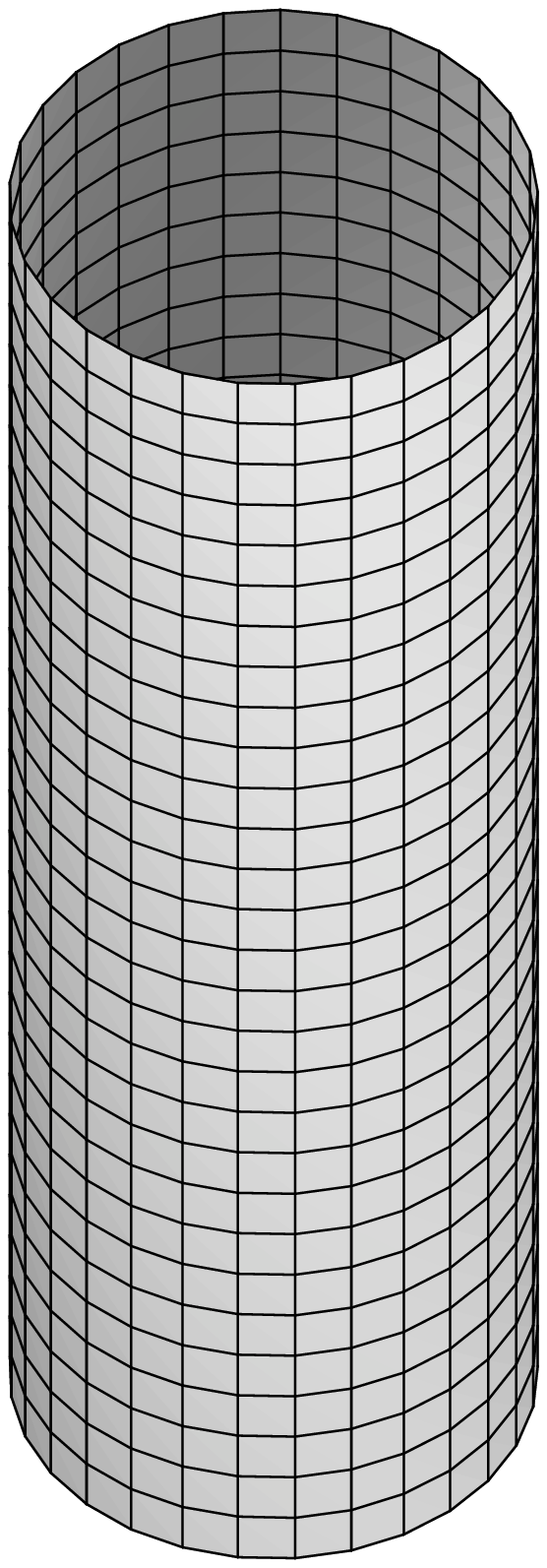}
\hspace*{2cm}
\includegraphics[width=2cm,height=4cm]{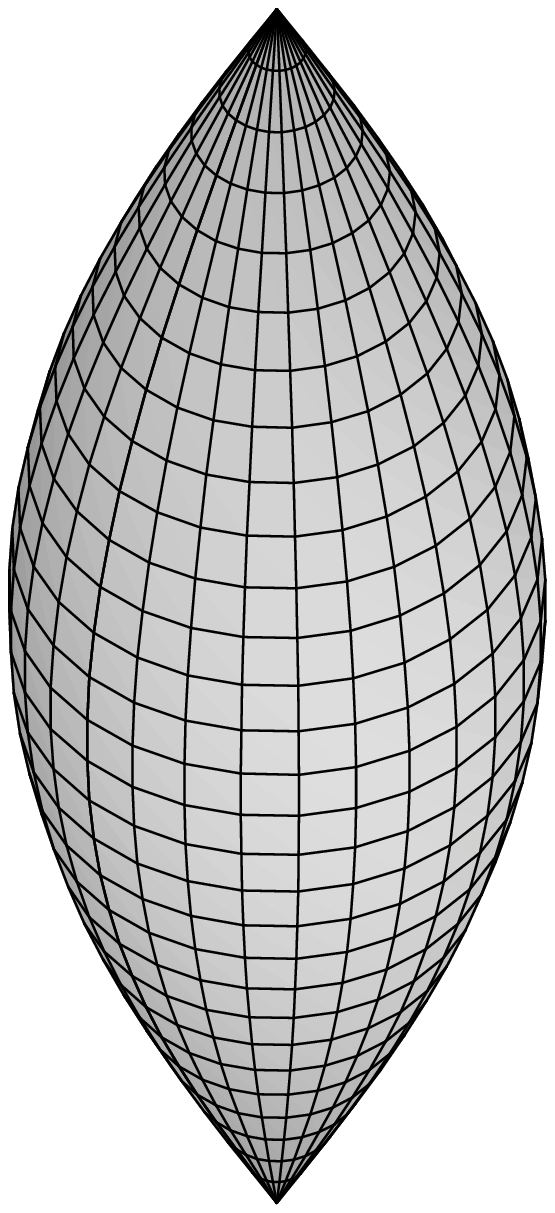}
\hspace*{2cm}
\includegraphics[width=2cm,height=4cm]{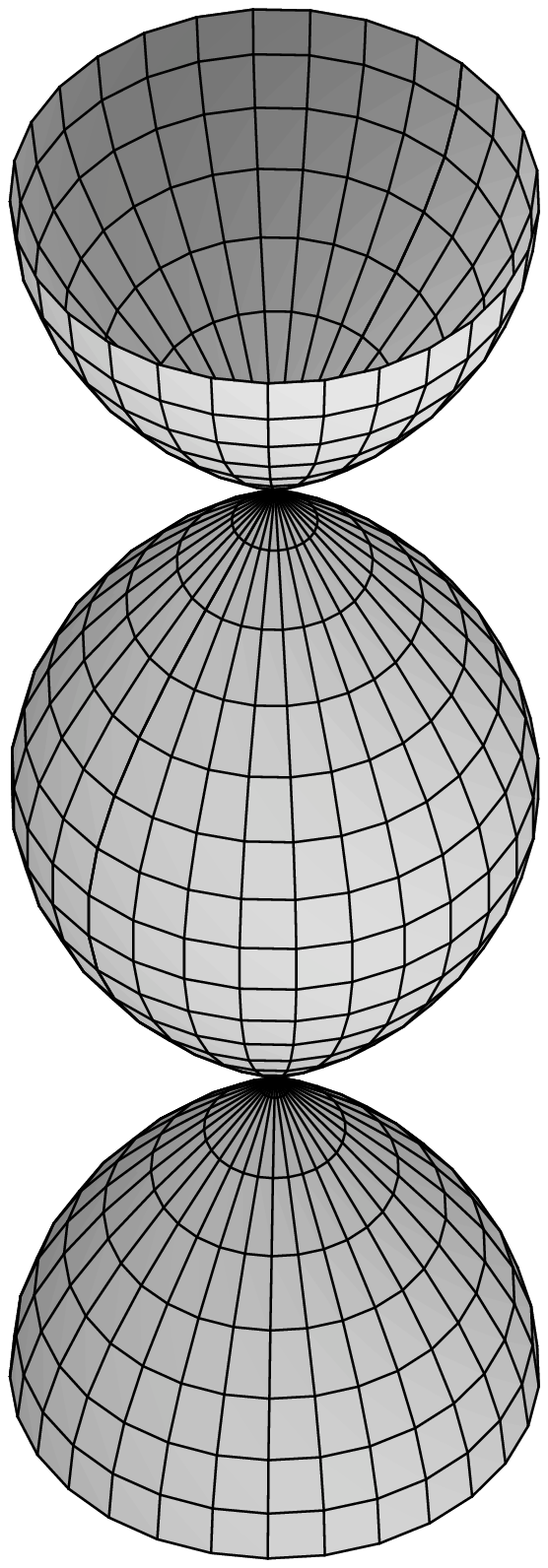}\\

p-wave\hspace*{33mm}$|{\bf \Delta}_2(\bf k)|$\hspace*{31mm}f-wave
\caption{The absolute value of the p-wave (left), ${\bf \Delta}_2(\bf k)$ (middle) and f-wave
 gap is shown. Note the presence of horizontal nodes in the latter two.\label{gaps}}
\end{figure}

However a simple analysis suggests that ${\bf \Delta}_2(\bf k)$ is
incompatible with the angular dependent magnetothermal conductivity data.
In fig. (\ref{gaps}) we show $\Delta(\bf k)$ for p-wave, f-wave and ${\bf \Delta}_2(\bf k)$
in quasi 2D systems.

We note here that ${\bf v}_c=0$ (i.e. the Fermi velocity in the $z$ direction)
 on the nodal lines of ${\bf \Delta}_2(\bf
k)$ which gives rise to a large $\cos(2\phi)$ term in the angular dependent
thermal conductivity\cite{dahm}.

In the following we compute the quasiparticle  density of states and the
optical conductivity for the f-wave gap and for the multigap model in the
Born limit at $T=0$K,
assuming, that both inter- and intraband scattering of impurities are
important,
 which was predicted to have strong effect on the
superconducting state\cite{agterberg}.
The density of states differs significantly in these two models, producing
distinct peaks from the active and passive bands in the ZR model, while
only one single peak is found in the f-wave case.
In the optical conductivity, beyond the intraband resonance peaks another
extra feature is found coming from interband scattering. On the other hand,
the single broad bump in the f-wave case can clearly be distinguished from
the three peak structure of ZR's model, hence both the single particle
density of states and the optical conductivity manifest evident signatures
of the corresponding model. 

\section{Density of states}

As already pointed out by Agterberg\cite{agterberg}, the effect of interband impurity
scattering plays an important role in the multigap model, similarly to the interband
scattering of Cooper pairs, which induced  ${\bf \Delta}_2(\bf k)$ in the
passive bands. However, in the f-wave case, we deal with a simple one band
model, therefore only intraband impurity scattering is considered.

\begin{figure}[h]
\psfrag{x}[t][b][1][0]{$\omega/\Delta_p$}
\psfrag{y}[b][t][1][0]{$N(\omega)/N_0$} 
\twofigures[width=7cm,height=6cm]{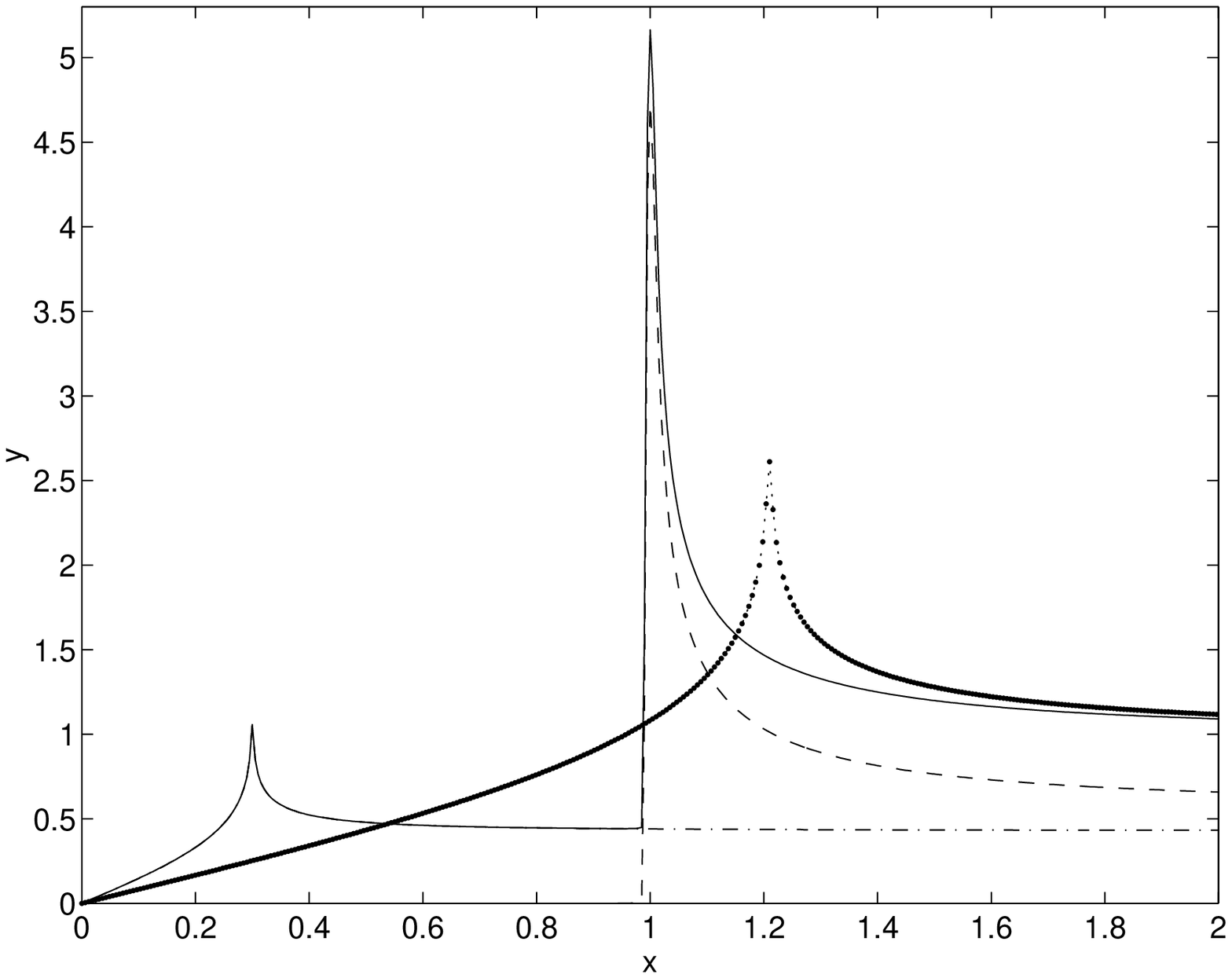}{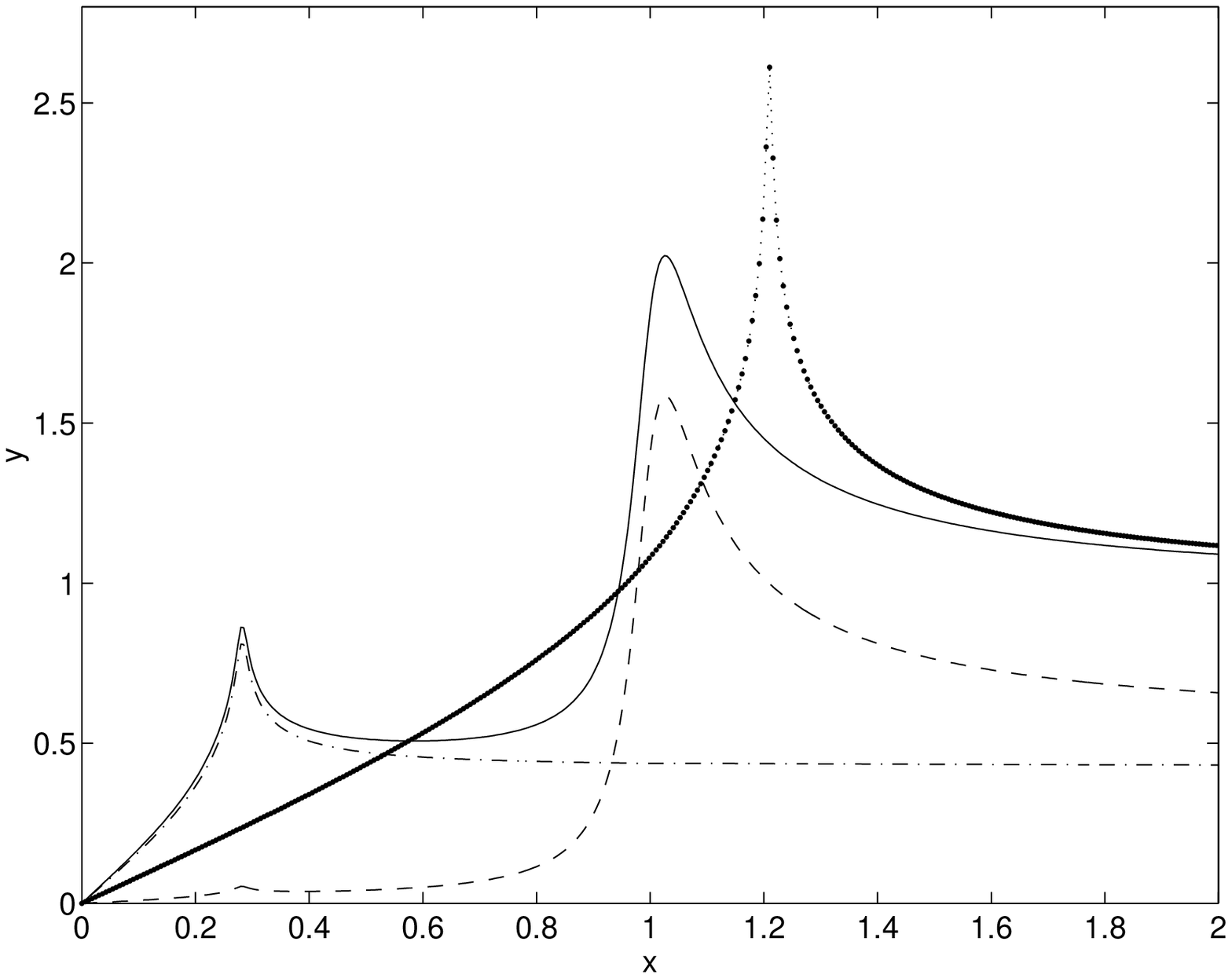}

\caption{The density of states is shown for weak interband scattering for the multigap model: in the
passive band (dashed-dotted line), active band (dashed line) and the full DOS
(solid line). The corresponding f-wave DOS (dotted line) exhibits one
single peak at $\omega=1.21\Delta_p$.\label{dos0p0010p0001}}
\caption{The density of states is plotted for strong interband scattering.
The peak in the p-wave channel (dashed line) at $\omega=0.3\Delta_p$
indicates the influence of these kind of processes.
\label{dos0p0010p1}}
\end{figure}

The effect of weak impurities (Born limit) is incorporated in the quasiparticle Green's
function of the multigap model by renormalizing the frequency:
\begin{gather}
\Delta_2
u_1=\omega+\Gamma_{11}\frac2\pi\frac{u_1}{\sqrt{1-u_1^2}}K\left(\frac{1}{\sqrt{1-u_1^2}}\right)+\Gamma_{12}\frac{u_2}{\sqrt{1-u_2^2}},\\
\Delta_p u_2=\omega+\Gamma_{21}\frac2\pi\frac{u_1}{\sqrt{1-u_1^2}}K\left(\frac{1}{\sqrt{1-u_1^2}}\right)+\Gamma_{22}\frac{u_2}{\sqrt{1-u_2^2}},
\end{gather}
where $u_1$ and $u_2$ are the dimensionless frequencies renormalized by
$\Delta_2$ and $\Delta_p$ in the passive
($\Delta_2$) and
active ($\Delta_p$) channels, respectively, $\Gamma_{jl}=n_i\pi N_l
|U_{jl}|^2$ ($j,l=1,2$), where the total density of states is split as
$N_1:N_2=0.43:0.57$ 
between the passive $\alpha+\beta$ and the active $\gamma$ bands from de
Haas-van Alphen measurement\cite{mackenzie}. $U_{11}$
and $U_{22}$ are the intraband, $U_{12}=U_{21}$ are the interband
scattering matrix elements, $n_i$ is the impurity concentration, $K(z)$ is
the complete elliptic integral of the first kind.
As opposed to this, for the f-wave model, the frequency is renormalized as
\begin{equation}
\frac{\omega}{\Delta_f}=u-\frac{\Gamma}{\Delta_f}\frac2\pi
\frac{u}{\sqrt{1-u^2}}K\left(\frac{1}{\sqrt{1-u^2}}\right),
\end{equation}
where $\Gamma=n_i\pi N_0 |U|^2$, $N_0=N_1+N_2$ is the total density of states.
At the Fermi energy ($\omega=0$), the frequencies are obtained as
\begin{subequations}
\begin{gather}
u_1=4i\exp\left(\frac\pi 2 \Delta_2\frac{\Delta_p-\Gamma_{22}}{\Gamma_{11}(\Delta_p-\Gamma_{22})+\Gamma_{12}\Gamma_{21}}\right),\\
u_2=\frac{\Delta_2}{\Delta_p}\frac{\Gamma_{21}}{\Gamma_{11}(1-\Gamma_{22})+\Gamma_{12}\Gamma_{21}}u_1,\\
u=4i\exp\left(\frac{\pi\Delta_f}{2\Gamma}\right)
\end{gather}
\label{zerusborn}
\end{subequations}
for small $\Gamma$'s.
The quasiparticle density of states (DOS) is obtained as
\begin{gather}
N_{\textmd{multigap}}(\omega)=\textmd{Im}\left(N_1\frac{u_1}{\sqrt{1-u_1^2}}K\left(\frac{1}{\sqrt{1-u_1^2}}\right)+N_2\frac{u_2}{\sqrt{1-u_2^2}}\right),\\
N_f(\omega)=N_0\textmd{Im}\frac{u_1}{\sqrt{1-u_1^2}}K\left(\frac{1}{\sqrt{1-u_1^2}}\right)
\end{gather}
for the multigap and f-wave model, respectively.
In the Born limit, the importance of interband scattering readily follows
from eq. (\ref{zerusborn}). Without interband processes, the two bands
decouple, and the p-wave gap behaves like an s-wave gap in the presence of
nonmagnetic impurities: it remains gapped for small impurity
concentrations, and reaches the gapless region only for high
concentrations or strong scatterers. In the density of states, peaks are expected at the gap
maxima of the two bands, namely at $\omega=\Delta_2$ and $\Delta_p$.
As soon as interband scattering is present, the p-wave channel becomes
gapless for arbitrary concentration and strength, as is seen from eq. \ref{zerusborn}.
In figs. \ref{dos0p0010p0001} and \ref{dos0p0010p1}, the density of states is
plotted for $U_{11}=U_{22}=U$, $\Gamma=0.001\Delta_p$ and for
weak ($\Gamma_{12}N_0/N_2=0.0001\Delta_p$) and strong ($\Gamma_{12}N_0/N_2=0.1\Delta_p$) 
interband scattering. Here we used $\Delta_2=0.3\Delta_p$ following refs. \cite{ZR,kusunose}, which was found
to describe very well the experimentally obtained specific heat and
penetration depth. Also for the f-wave model, the weak-coupling relation
$\Delta_f=1.21\Delta_p$ was assumed. In the weak interband scattering limit, 
the DOS in the $\gamma$ band increases very slowly, while for strong
impurities, even a small peak develops at $\omega=\Delta_2$ in the p-wave
DOS, due to the strong interplay of the two bands. 
For the f-wave model, only one peak is present at $\omega=\Delta_f$,
henceforth clear indications of either the multigap or the f-wave model
should be observed experimentally.

\begin{figure}[h]
\psfrag{x}[t][b][1][0]{$\omega/\Delta_p$}
\psfrag{y}[b][t][1][0]{$N(\omega)/N_0$} 
\twofigures[width=7cm,height=6cm]{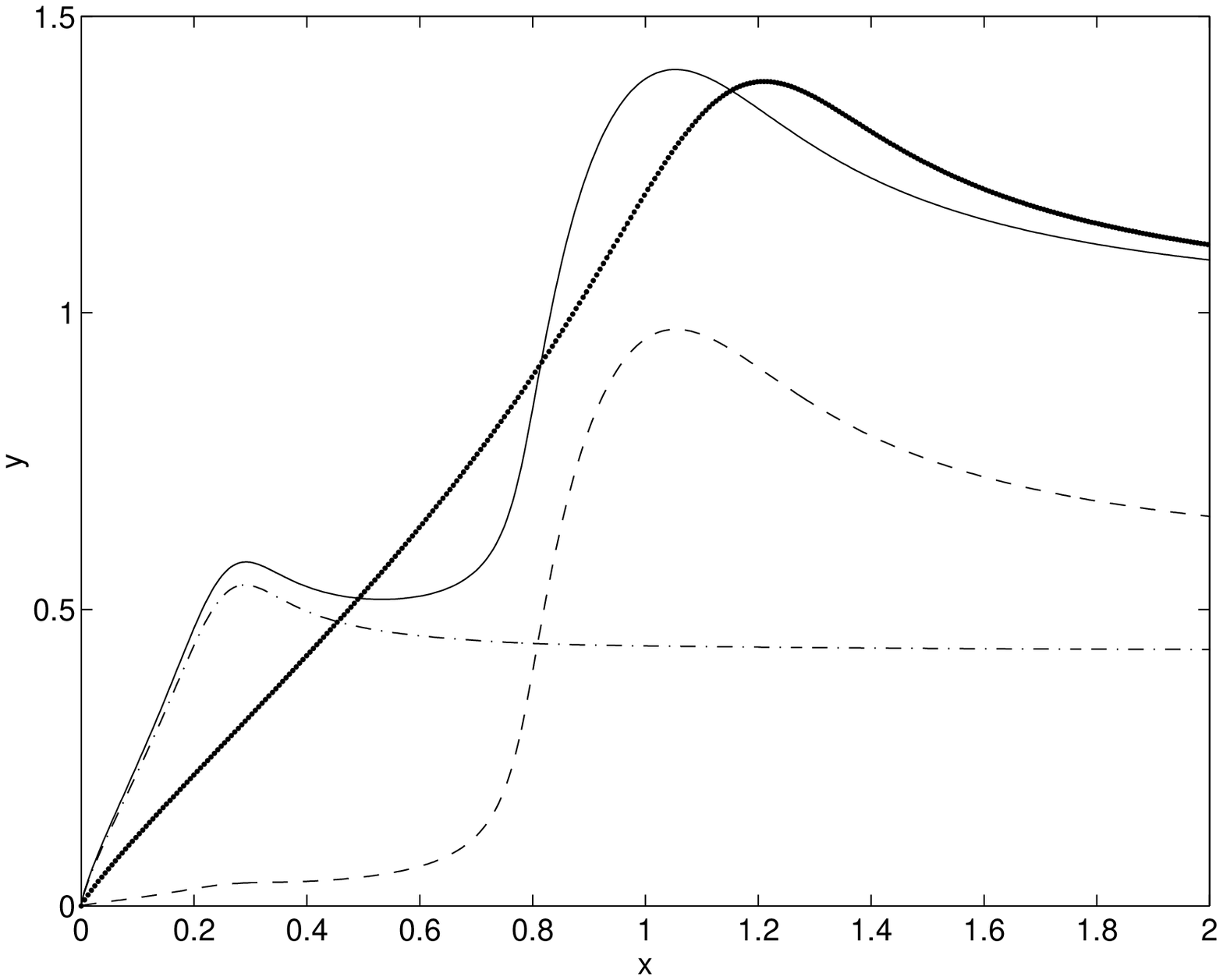}{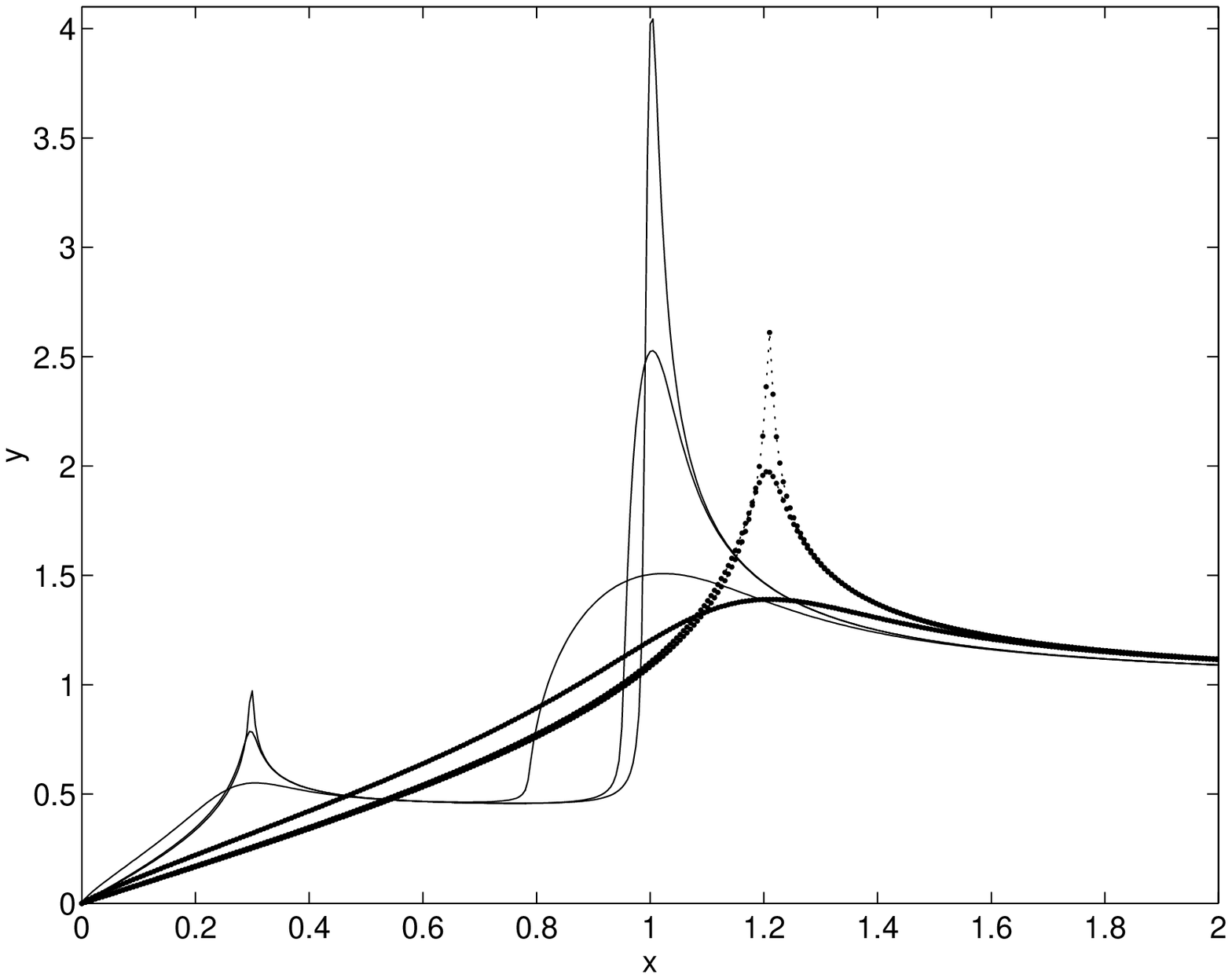}

\caption{The density of states is shown for medium strength impurities for
the multigap model: in the
passive band (dashed-dotted line), active band (dashed line) and the full DOS
(solid line). The corresponding f-wave DOS (dotted line) exhibits one
single peak at $\omega=1.21\Delta_p$.\label{dos0p1}}
\caption{The density of states is shown for the multigap (solid line) and
f-wave model (dotted line) for $U_{11}=U_{22}=U$, 
$\Gamma=0.001\Delta_p$, $0.01\Delta_p$ and
$0.1\Delta_p$, $\Gamma_{12}N_0/N_2$ is fixed to $0.01\Delta_p$.\label{dosx0p01}}
\end{figure}

In fig. \ref{dos0p1}, the DOS is shown for medium strength inter- and
intraband scatterers: 
 $U_{jl}=U$, $\Gamma=0.1\Delta_p$. 
The peaks at the gap maxima are broadened into big bumps and increasing density of states
is observable in the p-wave channel below the gap maximum. 
In fig. \ref{dosx0p01}, the DOS of both models are compared, as intraband 
scattering increases.

\section{Optical conductivity}

In the following we consider the optical conductivity of the multigap 
and f-wave model at $T=0$K. In the former case, as a lowest order approximation, each band contributes
separately to the optical response, which is obtained as
\begin{equation}
\textmd{Re}\sigma(\omega)=\zeta_1\textmd{Re}\sigma_1(\omega)+\zeta_2\textmd{Re}\sigma_2(\omega),
\end{equation}
where $\zeta_1=(n_1/m_1)/(n/m)=0.64$, $\zeta_2=(n_2/m_2)/(n/m)=0.36$, which measures the contribution
of each band to the conductivity\cite{mackenzie}.
At $T=0$K, the conductivities are obtained as
\begin{gather}
\textmd{Re}\sigma_l(\omega)=\frac{e^2n}{2\omega
m}\int\limits_0^\omega\textmd{Re}\left\{F_l[u_l(\omega-x),-u_l(x)]-F_l[u_l(\omega-x),-\overline{u_l(x)}]\right\}dx,
\end{gather}
where\cite{pdora,ddora}
\begin{gather}
F_1(u,u^\prime)=\frac{2}{\pi\Delta_2}\frac{1}{u^\prime-u}\left(\frac{u^\prime}{\sqrt{1-{u^\prime}^2}}K\left(\frac{1}{\sqrt
{1-{u^\prime}^2}}\right)-\frac{u}{\sqrt{1-u^2}}K\left(\frac{1}{\sqrt{1-u^2}}\right)\right),\\
F_2(u,u^\prime)=\frac{1}{\Delta_p}\frac{1}{u^\prime-u}\left(\frac{u^\prime}{\sqrt{1-{u^\prime}^2}}-\frac{u}{\sqrt{1-u^2}}\right).
\end{gather}
For the f-wave model, the conductivity reads as
\begin{equation}
\textmd{Re}\sigma(\omega)=\frac{e^2n}{2\omega
m}\int\limits_0^\omega\textmd{Re}\left\{F_1[u(\omega-x),-u(x)]-F_1[u(\omega-x),-\overline{u(x)}]\right\}dx,
\end{equation}
and $\Delta_2$ in $F_1(u,u^\prime)$ has to be replaced by $\Delta_f$.

\begin{figure}[h!]
\psfrag{x}[t][b][1][0]{$\omega/\Delta_p$}
\psfrag{y}[b][t][1][0]{Re$\sigma(\omega)2m\Delta_p/e^2m$} 
\twofigures[width=7cm,height=6cm]{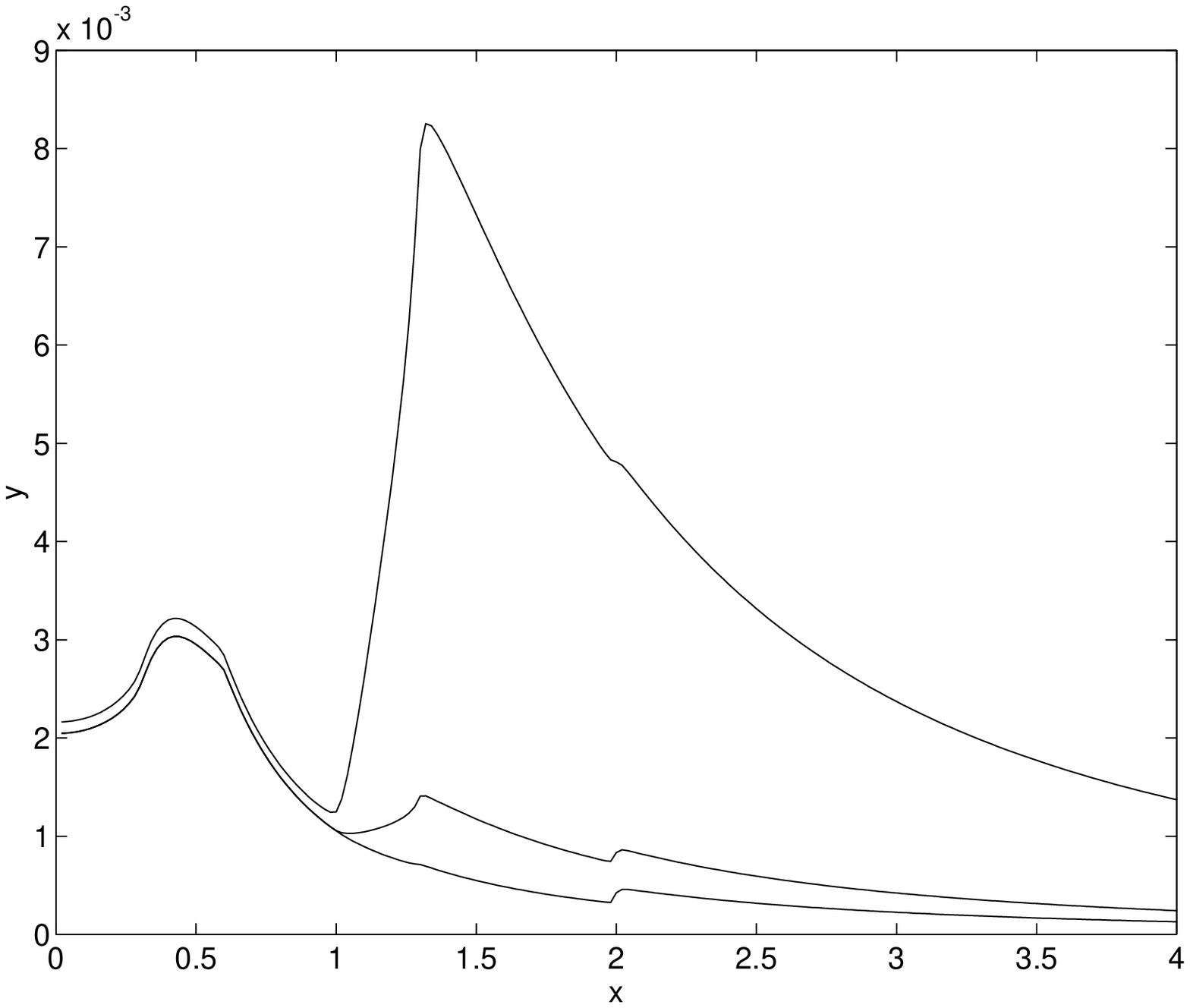}{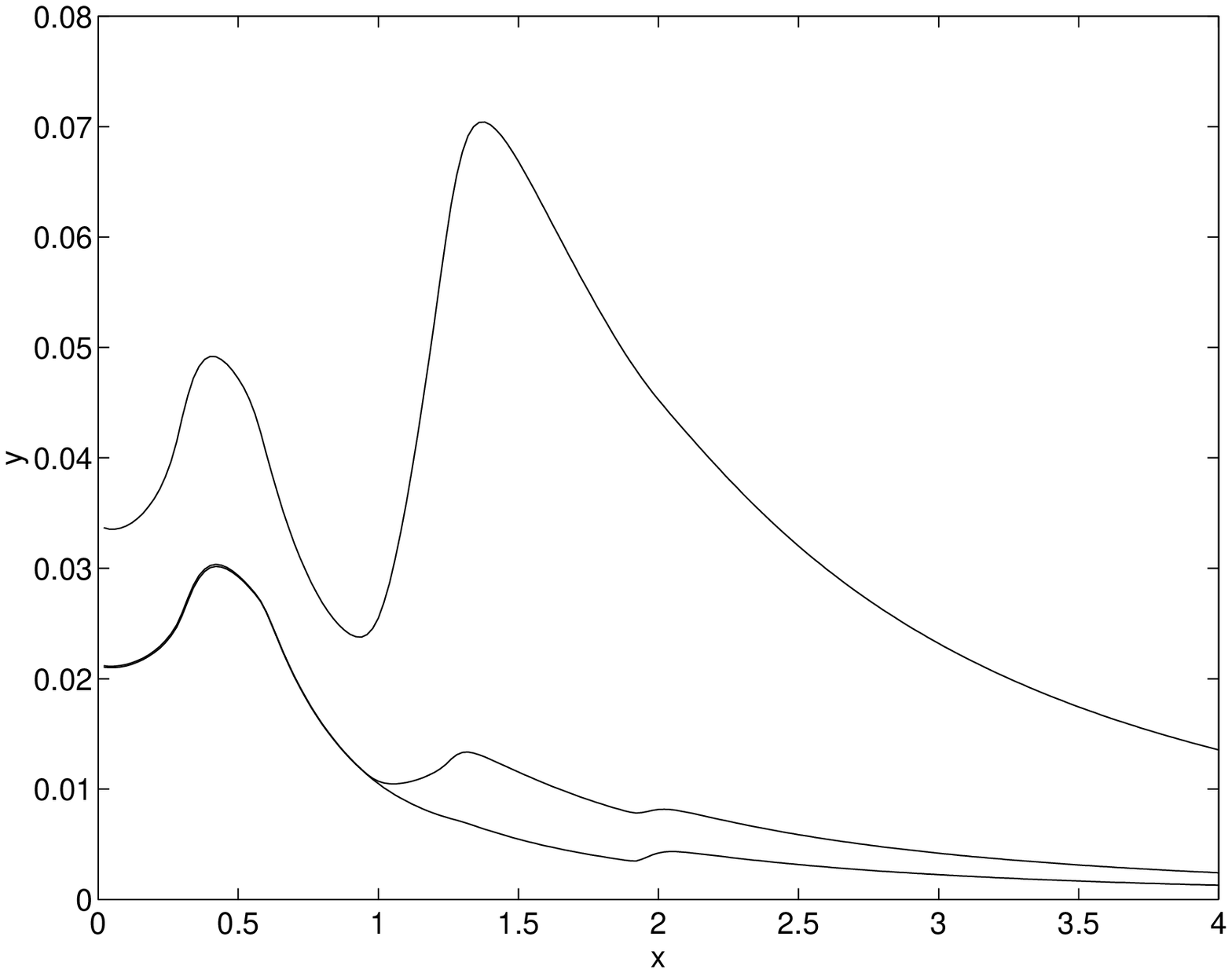}

\caption{The optical conductivity of the multigap model is plotted for
$U_{11}=U_{22}$, $\Gamma_{11}N_0/N_1=0.001\Delta_p$, the
interband scattering parameter changes as $\Gamma_{21}=0.1\Gamma_{11}$, 
$\Gamma_{11}$ and $10\Gamma_{11}$, with increasing peak at $\omega=1.3\Delta_p$.\label{vmb0p001}}
\caption{The optical conductivity of the multigap model is plotted for
$U_{11}=U_{22}$, $\Gamma_{11}N_0/N_1=0.01\Delta_p$, the
interband scattering parameter changes as $\Gamma_{21}=0.1\Gamma_{11}$, 
$\Gamma_{11}$ and $10\Gamma_{11}$,  with increasing peak at $\omega=1.3\Delta_p$.
\label{vmb0p01}}
\end{figure}

\begin{figure}[h!]
\psfrag{x}[t][b][1][0]{$\omega/\Delta_p$}
\psfrag{y}[b][t][1][0]{Re$\sigma(\omega)2m\Delta_p/e^2n$} 
\twofigures[width=7cm,height=6cm]{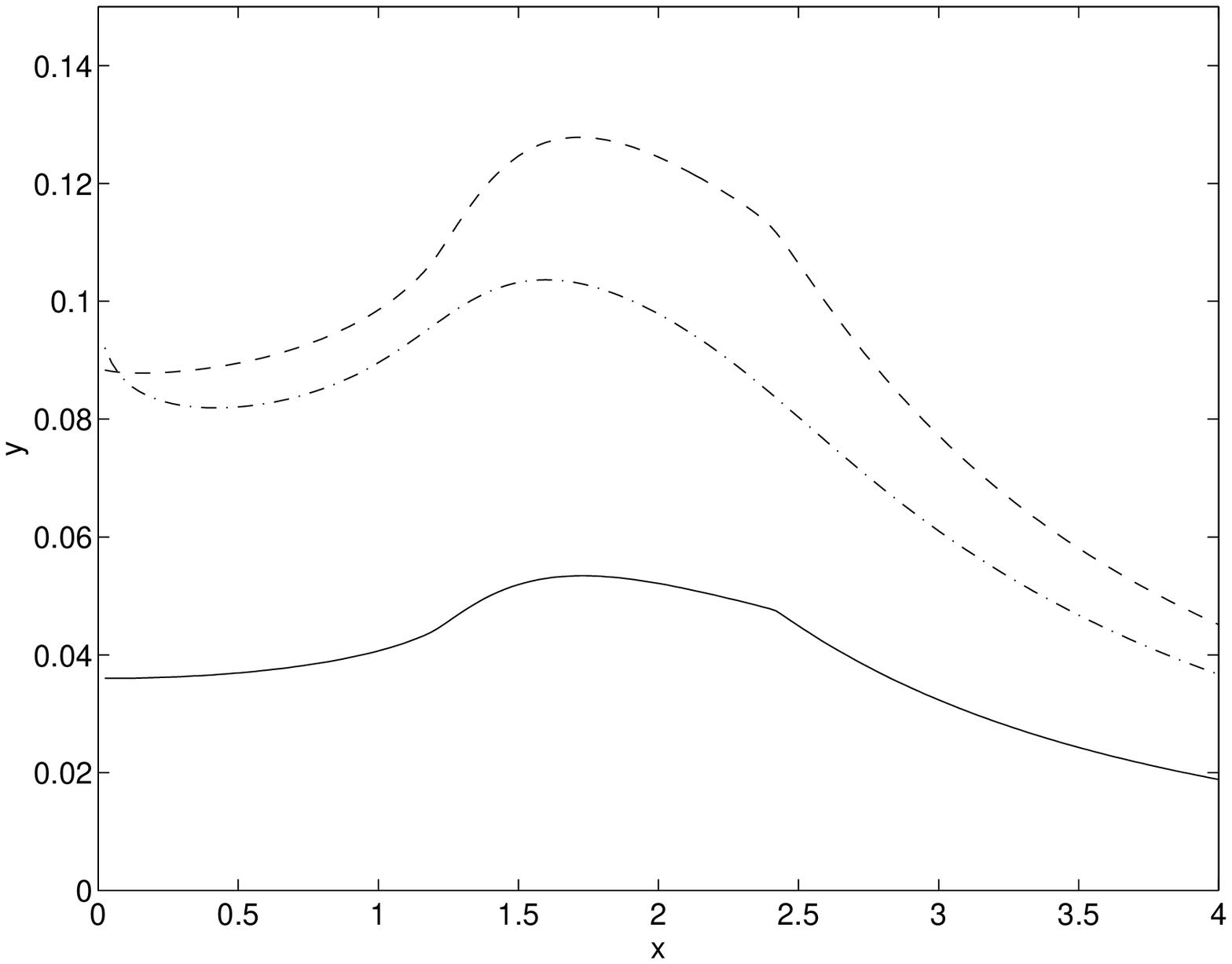}{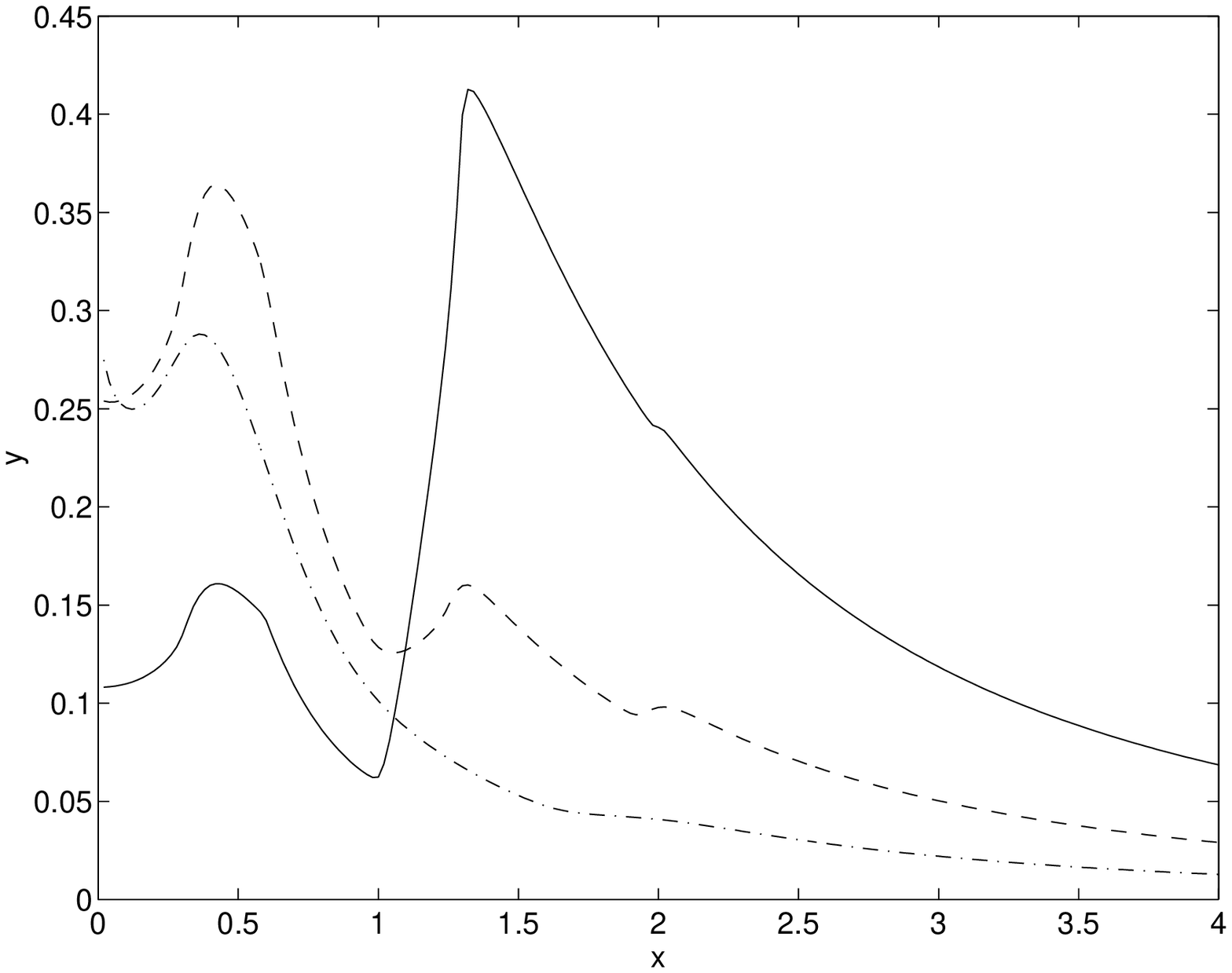}

\caption{The optical conductivity of the f-wave model is shown for
$\Gamma_{11}N_0/N_1=0.001\Delta_p$
(solid line, enlarged by 50), $0.01\Delta_p$ (dashed line, enlarged by 12) and $0.1\Delta_p$ (dashed-dotted
line).\label{vfb}}
\caption{The optical conductivity of the multigap model is shown for
$\Gamma_{12}N_0/N_2=0.01\Delta_p$, $\Gamma_{11}N_0/N_1=0.001\Delta_p$
(solid line, enlarged by 50), $0.01\Delta_p$ (dashed line, enlarged by 12) and $0.1\Delta_p$ (dashed-dotted
line).\label{vmbx0p01}}
\end{figure}

>From the density of states of the multigap model, one would expect peaks in
the conductivity at $\omega=2\Delta_2$, $2\Delta_p$, and if interband
scattering is strong enough, at $\omega=\Delta_p+\Delta_2$. These
features can bee seen in figs. \ref{vmb0p001} and \ref{vmb0p01}. Here the
intraband scattering parameter was fixed as
$U_{11}=U_{22}$, $\Gamma_{11}N_0/N_1=0.001\Delta_p$ and
$0.01\Delta_p$, respectively, and the
interband scattering parameter varied from weak through medium to
strong as $\Gamma_{21}=0.1\Gamma_{11}$, $\Gamma_{11}$ and $10\Gamma_{11}$. 
Along this line, as $\Gamma_{21}$
enhances, the peak corresponding to interband
excitations at $\omega=\Delta_p+\Delta_2$ becomes dominant, overwhelming
the intraband processes.

In the f-wave case, the conductivity is identical to those of a d-wave
superconductor, and exhibits a small peak at $\omega=0$ as the impurities
get stronger, and at the same time a rather broad bump is developed at
$\omega=1.5-2\Delta_p$ as is seen in fig. \ref{vfb}.
It is remarkable also, that in spite of the different scattering rates
changing within two orders of magnitude, the
curves look rather similar to each other, and can easily be distinguished
from the multigap model.
For the multigap model, as intraband scattering becomes dominant, the
conductivity of the passive band rules over the other transitions, since the
same impurities decrease the superfluid density much faster in this band
due to the smaller gap (see fig. \ref{vmbx0p01}!).

\section{Conclusion}

We have calculated the density of states and the optical conductivity for ZR's multigap model and for
the f-wave case in the Born scattering limit. The effect of interband
scattering produces new structures in the optical response of the multigap
model: resonant peaks can be observed at $\omega=2\Delta_2$, $2\Delta_p$, and due
to interband transitions at $\Delta_2+\Delta_p$.
Therefore when the experimental difficulty to perform the optical  measurement
below $T=0.1$K is overcome, this will provide a unique way to explore the
validity of models for the superconductivity in Sr$_2$RuO$_4$. In a
forthcoming paper we shall discuss the effect of resonant scatterers, where
we also expect clear differences between the multigap and f-wave model.

\section{Acknowledgments}
We are
benefited from useful discussions with Hyekyung Won and Peter Thalmeier. 
This work  
was supported by the Hungarian National Research Fund under grant numbers
OTKA T032162 and T037451.

\bibliographystyle{apsrev}
\bibliography{srruo}
\nocite{*}

\end{document}